\documentclass [10pt]{article}
\usepackage{amsmath}
\usepackage{epsfig}

\begin{document}
\baselineskip=15pt \hsize=340pt \vsize=490pt
\def\b{\bar}
\def\d{\partial}
\def\D{\Delta}
\def\cD{{\cal D}}
\def\cK{{\cal K}}
\def\f{\varphi}
\def\g{\gamma}
\def\G{\Gamma}
\def\l{\lambda}
\def\L{\Lambda}
\def\M{{\Cal M}}
\def\m{\mu}
\def\n{\nu}
\def\p{\psi}
\def\q{\b q}
\def\r{\rho}
\def\t{\tau}
\def\x{\phi}
\def\X{\~\xi}
\def\~{\widetilde}
\def\h{\eta}
\def\bZ{\bar Z}
\def\cY{\bar Y}
\def\bY3{\bar Y_{,3}}
\def\Y3{Y_{,3}}
\def\z{\zeta}
\def\Z{{\b\zeta}}
\def\Y{{\bar Y}}
\def\cZ{{\bar Z}}
\def\`{\dot}
\def\be{\begin{equation}}
\def\ee{\end{equation}}
\def\bea{\begin{eqnarray}}
\def\eea{\end{eqnarray}}
\def\half{\frac{1}{2}}
\def\fn{\footnote}
\def\bh{black hole \ }
\def\cL{{\cal L}}
\def\cH{{\cal H}}
\def\cF{{\cal F}}
\def\cP{{\cal P}}
\def\cM{{\cal M}}
\def\ik{ik}
\def\mn{{\mu\nu}}
\def\a{\alpha}

\title{Stringlike structures in Kerr-Schild geometry: N=2 string, twistors and
 Calabi-Yau twofold}
\author{Alexander Burinskii}

\author{Alexander Burinskii, \\
Theor. Phys. Lab., NSI, Russsian Academy of Sciences,\\
B. Tulskaya 52,  Moscow 115191 Russia}

\maketitle
\begin{abstract}
\noindent The four-dimensional
Kerr-Schild geometry contains two stringy structures. The
first one is the closed string formed by the Kerr singular ring, and the
second one is an open complex string with was obtained in the complex
structure of the Kerr-Schild geometry. The real and complex Kerr
 strings form together a membrane source of the over-rotating
 Kerr-Newman solution without horizon, $a =J/m >> m .$
It has also been obtained recently that the principal null congruence of
the Kerr geometry, induced by the complex Kerr string, is determined by the Kerr theorem
as a quartic in the projective twistor space, which corresponds
to embedding of the Calabi-Yau twofold in the bulk of the Kerr geometry.
In this paper we describe this embedding in details and  show that the
four folds of the twistorial K3 surface represent an analytic extension of the
Kerr congruence created by antipodal involution.
\end{abstract}


\maketitle

\section{Introduction}
It is now commonly accepted that black holes (BH) are to be associated with elementary particles. Physics of black holes is based on the complex analyticity and conformal field theory, which unites them with superstring theory and particle physics.
In the previous paper \cite{BurAlter} (hereafter referred as I), we considered emergence of the string-like structures in the four-dimensional Kerr-Schild geometry without horizons.  In the dimensionless units $G=c=\hbar =1 ,$ this happens when the Kerr rotational parameter $a=J/m$  exceeds the mass parameter $m .$ This case of the over-rotating Kerr geometry is very important for application to elementary particles. So far as masses of the elementary particles are very small, and the spin may be extreme high, one obtains that for spinning particles $a$ exceeds $m$ for
many orders. In particular,  mass of an electron in the dimensionless units is
$m\sim 10^{-22} ,$ while spin $J = \hbar /2 =1/2 .$ It yields $a=J/m \sim 10^{22} ,$ and therefore, the value $a$ exceeds $m$  for about $44$ orders.

Principal peculiarity of the over-rotating Kerr geometry is presence of the naked Kerr singular ring, which is branch line of the Kerr space into two sheets, and therefore,  it forms a stringlike topological defect of the space-time. The singular metric should be regularized to an almost flat space-time to get a correspondence with quantum theory and the experimentally negligible role of gravity in particle physics. This procedure was specified step by step by many authors during about four decades, and resulted in formation of the smooth and regular source of the Kerr-Newman solution in the form of a relativistically rotating and highly oblate vacuum bubble.  Structure of this source and the corresponding  references were discussed in I and \cite{BurSol}.

The region near the Kerr singular ring is excised and replaced by a vacuum bubble, while the Kerr-Newman electromagnetic field is regularized and gets a finite maximal
value at the edge rim of the oblate bubble, forming a stringlike loop around the bubble. The very old proposal
that the Kerr singular ring plays the role of a closed string \cite{IvBur} was later supported by the systematic investigations of the singular string solutions of the low energy string theory as solitons \cite{DGHW}. It has been shown in \cite{BurSen} that the structure of fields around the  Kerr singular
   ring is similar to structure of the fundamental  string solutions  obtained by Sen \cite{Sen,KerSen}.

One more  stringy structure was obtained in the complex representation of the Kerr geometry, in which the Kerr solution is represented as a retarded-time field from  the particle propagating along a complex world line \cite{LinNew,BurKerr,BurNst}. However, the  complex world line is really a world sheet \cite{OogVaf}, and therefore, it is equivalent to a string \cite{BurCStr,BurTwi}. Complex world sheet of his string is embedded in the real Kerr geometry, and as it discussed in I, the real and complex Kerr strings form together a membrane, which is analogous of the enhancon model of the string/M-theory unification, \cite{LaJoh}.
 Recently, the both Kerr strings were mentioned in \cite{AdNew} by Adamo and Newman, and they reaction is worth quoting: ``...It would have
been a cruel god to have layed down such a pretty scheme and not have it mean something deep.''

In addition to this stringy system, one  more remarkable fact  was obtained in the paper I,  emergence of the Calabi-Yau twofold
(K3 surface) in the  projective twistor space of the Kerr geometry. It appears
 as a quartic  on the projective twistor space $CP^3 ,$ and determines the Kerr Principal Null Congruence (PNC) in accordance with  the Kerr theorem.
 In paper I we arrive at the conclusion, that this parallelism is not accidental, and there may be
a fundamental underlying structure lying beyond these relationships. In particular, it was supposed that
this correspondence may be related with a mysterious complex N=2 string,  which has critical real dimension four \cite{GSW} and is close related with twistors \cite{OogVaf,Gibb}.  In this paper we extend and specify the given in I
treatment of the Calabi-Yau space K3 embedded into four-dimensional Kerr-Schild (KS) geometry
and give extra arguments in favor of the assumption that the source of the striking parallelism with superstring theory is the N=2 critical superstring, structure of which is consistent with twistorial structure of KS geometry.
We work in the Kerr-Schild formalism \cite{DKS}, which is parallel to the Newman-Penrose formalism, but has the advantage, related with action of the Kerr theorem, which allows to describe the geodesic and shear-free congruences (GSF)  on the auxiliary Minkowski background in terms of twistors.

\section{Basics of the Kerr-Schild formalism}
{\bf Structure of the Kerr-Newman solution.}
 KN metric is represented in the Kerr-Schild (KS) form \cite{DKS},
 \be g_\mn=\eta _\mn + 2h
e^3_\m e^3_\n \label{KSh} , \ee where $\eta_\mn$ is auxiliary
Minkowski background in Cartesian coordinates ${\rm x}= x^\m
=(t,x,y,z)$ with signature $(-+++) ,$ and \be h = P^2 \frac {mr-e^2/2} {r^2 + a^2 \cos^2
\theta}, \quad P=(1+Y\Y)/ \sqrt 2 ,  \label{h}\ee and $e^3 (\rm
x)$ is a tangent direction to a \emph{Principal Null Congruence
(PNC)}, which is determined by the form \be e^3_\m
dx^\m =du + \bar Y d \zeta + Y d \bar\zeta - Y\bar Y dv ,
\label{e3} \ee
where \be \z = (x+iy)/\sqrt 2 ,\quad  \Z = (x-iy)/\sqrt 2 , \quad
u = (z + t)/\sqrt 2 ,\quad v = (z - t)/\sqrt 2 \label{nullcoord}\ee are the null Cartesian coordinates, and
function $Y ( x^\m)$  is obtained from
\emph{the Kerr theorem},
\cite{KraSte,Pen,PenRin,BurKerr,Multiks,MultPart,BurNst}.
Kerr's oblate spheroidal coordinates $r, \theta, \phi$ are related to Cartesian coordinates as follows
\be x+iy  = (r + ia) e^{i\phi} \sin \theta,  \qquad
z = r\cos\theta, \qquad \rho = t - r . \label{oblate} \ee
The Kerr singular ring corresponds to caustic of  PNC  at
$r=\cos\theta=0 . $  One sees that the metric (\ref{KSh}) and
electromagnetic potential of the the KN solution, \be A_{\m} =
-P^{-2} Re \frac {e} { (r+ia \cos \theta)} e^3_\m \label{Amu} ,
\ee are aligned with directions $e^3_\m$ of the Kerr PNC.
\noindent By regularization of the KN solution, the Kerr singular
ring is excised, forming a vacuum bubble, boundary  of which is
determined by the condition $h=0 $ corresponding to Minkowski
metric inside  the bubble and at the boundary. From (\ref{h}) one
sees that this boundary corresponds to  $r=r_e = e^2/2m ,$ and (\ref{oblate}) shows
that it forms a highly oblate ellipsoidal membrane. The KN electromagnetic field
is concentrated at the sharp edge of the membrane and regularized by the
cut-off at $r=r_e .$

\paragraph{Kerr theorem.}
Kerr theorem determines the shear free null congruences
\cite{DKS,KraSte} with tangent direction (\ref{e3}) via a complex
function $Y(x^\m),$ which is solution of the equation \be F(T^A) =0
\label{F0KerrTeor}, \ee where $F(T^A)$ is any
holomorphic function of the projective twistor coordinates   \be
T^A= \{ Y, \quad \l ^1 = \z - Y v, \quad \l ^2 =u + Y \Z \}, \quad A=1,2,3
.\label{TA} \ee Using the  Cartesian coordinates $x^\m ,$ one
can rearrange coordinates and reduce function $F(T^A)$ to the form
$F(Y,x^\m),$ which allows one to get explicit solution of the
equation (\ref{F0KerrTeor}) in the form $Y(x^\m).$

\noindent For the Kerr and KN solutions function $ F(Y,x^\m)$
is quadratic in $Y,$ \be  F = A(x^\m)  Y^2 + B(x^\m)
Y + C(x^\m), \label{FKN} \ee and the equation (\ref{F0KerrTeor})
represents
 a \emph{quadric }in the projective twistor space $\bf CP^3 .$
If determinant $\D = (B^2 - 4AC) $ is non-degenerate, it
corresponds to the Kerr complex radial distance \cite{BurKerr,BurNst}
\be \tilde r \equiv r+ia \cos \theta= - \D^{1/2} , \label{trDet} \ee
 and the solutions of (\ref{F0KerrTeor}), \be Y^\pm (x^\m)= (- B \mp \tilde r )/2A, \label{Ypm}\ee
determine two different Kerr's congruences via expression (\ref{e3}).
Another expression for the complex radial distance, \be \tilde r =
- dF/dY \label{tr} ,\ee may be obtained from the relations
(\ref{FKN}) and (\ref{Ypm}).  It shows that the Kerr singular ring $\tilde r =0$ is caustic of the Kerr congruence,  $ dF/dY=0 .$

As a consequence of the Vieta's formulas, the quadratic in $Y$
function (\ref{FKN}) may also be expressed via the solutions
$Y^\pm(x^\m)$ in the form \be
F(Y,x^\m)=A(Y-Y^+(x^\m))(Y-Y^-(x^\m)) \label{FYYpm} .\ee

\begin{figure}[ht]
\centerline{\epsfig{figure=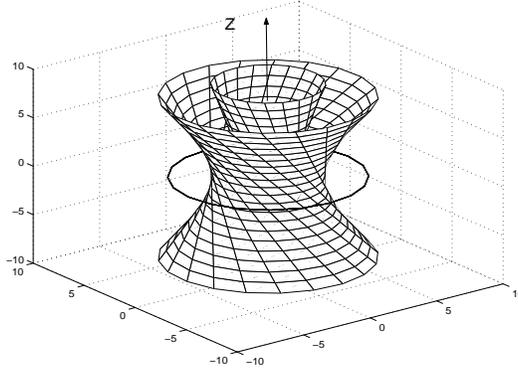,height=5cm,width=7cm}}
\caption{Kerr's congruence of twistor null lines (PNC) is focused on
the Kerr singular ring, forming a branch line of the KS space into two sheets. }
\end{figure}

\noindent{\bf Stationary KS geometries}
The Kerr and KN solutions  are stationary solutions of
the Einstein-Maxwell field equations. There are also more general stationary KS solutions for which
 \be A_{\m} = -P^{-2} Re \frac {\psi(Y)} { (r+ia \cos \theta)} e^3_\m \ee and \be h= P^2 \frac {mr- |\psi(Y)|^2/2} {r^2 + a^2 \cos^2 \theta} ,\ee where $\psi(Y)$ is arbitrary analytic function. The poles of $\psi(Y)$  correspond to singular beams supported by  twistor null lines of the Kerr congruence, \cite{BurA}. The stationary solutions  are characterized by a constant real Killing
direction $K^\m,$ which corresponds to invariance of the metric $g_\mn$ with
respect to the action of the Killing operator $ \hat K = K^\m \d_\m ,$
 \be\hat K g^\mn =0 . \label{station}\ee
 It imposes requirements on the Kerr
congruence, $\hat K e^3 =0 ,$ on  the functions
$Y^\pm(x)$ and $\bar Y^\pm(x),$ \be
\hat K Y =\hat K \Y =0 \label{KY} ,\ee and as a consequence,
on the Kerr generating function $ F(Y, x^\m)$ and on the coefficients
$\hat K A =\hat K B =\hat K C = 0 \label{KA} .$

\section{Complex structure of the Kerr geometry}
 The complex radial distance $\tilde r = r+ i a\cos\theta $ takes in
the Cartesian coordinates (\ref{oblate}) the form $ \tilde r =\sqrt {x^2+y^2
+(z+ia)^2},$
and therefore, the scalar component of the vector potential (\ref{Amu}) may be
obtained from the Coulomb solution $\phi(\vec x) = e/r = e/\sqrt{x^2+y^2
+z^2} $ by a complex shift $z\to z+ia ,$ or by  the shift of its singular origin
 $\vec x_L =(0,0,0)$ in the complex direction $\vec x_L \to (0,0, -ia).$
The complex shift was first considered  by Appel in 1887  \cite{App}, who noticed
that the Coulomb solution is invariant with respect to the real shift of coordinates,
and due to linearity of the Laplace equation, it should also be invariant with
respect to the complex shift of its origin, which is trivial gauge transformation.
This procedure results in nontrivial consequences for the real section of the solution.
In particular, singular point of the Coulomb solution $\vec x_L =(0,0,0)$ turns
into singular ring $x^2+y^2+(z+ia)^2 =0 $ (intersection of the sphere $x^2+y^2+z^2 =a^2 $
and plane $z=0 $), which becomes a branch line of the space creating a twosheeted space-time.

It has been shown \cite{LinNew,Bur0} that the KN electromagnetic field may be obtained by the complex shift from
the Coulomb solution. and Lind and Newman suggested in \cite{LinNew} a retarded-time construction, in which the linearized KN solution was represented as  field generated by a complex source propagating
along a complex world line (CWL). Later on, it was obtained in \cite{BurKerr,BurNst} that the retarded-time construction turns out to be exact representation in the Kerr-Schild formalism. This \emph{exactness} appears as a consequence of
the specific linearization of the KS gravity due to analyticity of the twistorial structure determined by the Kerr theorem.

Principal element of the complex retarded-time construction is a family of the
 complex light cones $\cK$ adjoined to each point of the CWL.
 Taking the Left CWL as a basis for treatment,
 one can represent the family of the adjoined complex light cones in spinor form
 \be {\cal K}_L = \{x: x = x^{i}_L(\tau_L) + \psi ^{A}_{L} \sigma ^{i}_{A \dot { A}}
 \tilde{\psi}^{\dot{A}}_{R} \} . \label{KL}  \ee
\noindent The cones are split  into
two families of null planes: "Left" $( \psi _{L}$ =const; $\tilde{\psi
}_{R}$ -var.) and "Right"$( \tilde{\psi }_{R}$ =const; $\psi _{L}$ -var.).
The null rays of the Kerr congruence appear as a real slice of the Left (or Right)
complex null planes on the KS auxiliary  Minkowski background.
The real null direction $e^3_\m , $ (\ref{e3}) is a
projective form of the spinor expression  $ \psi ^{A}_{L} \sigma_{\m A \dot { A}}
\bar {\psi}^{\dot{A}}_{L}  ,$ in which the spinor $\psi_L^A$ is replaced by the first
projective twistor coordinate $T^1 \equiv Y =\psi_L^2/\psi_L^1.$
The real form $e^3$ is completed by two complex conjugate null forms
\be e^1 = d \zeta - Y dv, \qquad  e^2 = d \bar\zeta -  \bar Y dv.
\label{e12} \ee
One sees that the second projective twistor coordinate
$T^2 \equiv \l ^1 =\z - Y v = x^\m e^1_\m $ represents
projection of the space-time point $x^\m$ on the null direction $e^1 ,$ while the third
 projective twistor coordinate $ T^3 \equiv \l ^2 =u
+ Y \Z = x^\m (e^3_\m- \Y e^1_\m) $ represents a linear combination of the projections on the
null directions $e^3$ and $e^1 .$ The determined by the Kerr theorem function $Y(x)$ allows one to restore at each point $x$ the remainder twistor coordinates $ \l ^1$ and $ \l ^2$, and
to fixe the incident Left null plane spanned by the null
directions $e^3$ and $e^1 .$ It should be mentioned \emph{analyticity} of this construction. The
Left null plane connects the real points $x^\m$ with a complex point of the Left CWL $x_L^\m.$ It belongs simultaneously to the ''in''-
fold of the complex light cone adjoined to the point $x^\m$ and to the
''out''-fold of the light cone emanating from a point of the Left CWL $x_L^\m.$
The point of intersection of the Left null plane with CWL $x_L^\m$ fixes the value of the
Left retarded time $\t_L$ , which becomes an analytic complex scalar
function on the real space-time $\t_L(x^\m).$ In \cite{BurNst}  this
incidence was called as L-projection.

\noindent Similar system of the  Right null planes, spanned by  null directions $e^3$ and $e^2 ,$
 connects  real points $x^\m$ with Right CWL and determines \emph{anti-analytic} R-projection and the retarded
times $\t_R (x^\m).$ For the real $x^\m$ functions $Y(x^\m)$ and $\Y(x^\m)$ are complex
conjugate and determine the conjugate Right and Left null planes. Analyticity of the generating function of the Kerr theorem $F(T^A)$ and the corresponding
solutions $Y^\pm(x^\m) $ creates independence of the analytic and anti-analytic L- and R- structures, allowing the complex-analytic extension of the real KS geometry.
In accord to analyticity of the Kerr theorem, the L-projection foliates the complexified Minkowski space-time $CM^4$ into the Left null planes.
Due to algebraically specifical structure of the KS metric (\ref{KSh}) and the gauge field
(\ref{Amu}), the non-linear terms drop out, and the analytic  Left foliation
performs a specific \emph{linearization of the KS background}. In particular,  it results in a \emph{twistorial version } of the Fourier transform (introduced by Witten in \cite{Wit}), which acts on the null planes of the curved KS space-time performing a bridge between KS gravity and quantum theory.

 The stationary KN solution is represented as a retarded-time field generated by a
{\it complex source moving along  a straight CWL}, \cite{BurKerr,BurNst},
\be x_L^\m(\t_L) = x_0^\m (0) +  u^\m \t_L + \frac{ia}{2} \{ k^\m_L -
k^\m_R \} ,\label{CWL}\ee where $u^\m=(1,\vec v)$ is a real
4-velocity, $ k_R=(1,0,0,1), \ k_L=(1,0,0,-1),$ and $\t_L=t_L+ i\sigma_L$ is a
complex retarded-time parameter. Index $L$ labels the
Left complex structure, and we should also add the complex conjugate Right one
\be x_R^\m(\t_R) = x_0^\m (0) +  u^\m \t_R  -
\frac{ia}{2} \{ k^\m_L - k^\m_R \} .\label{CWR}\ee  Complex shift turns
the hedgehog of the Schwarzschild radial directions $\vec n = \vec r /|r|$ into
twisted directions of the Kerr congruence,  Fig.1.

\section{Complex open string}
 The complex world line $x_0^\m (\t) ,$
parametrized by complex time $\t,$ represents a
two-dimensional surface which takes an intermediate position
between particles and strings \cite{BurCStr,OogVaf}.
The corresponding  "hyperbolic string" equation \cite{OogVaf},
 \be \d_\t \d_{\bar\t}
x_0(t,\sigma) =0 , \label{StrEq}\ee corresponds to bosonic part of the complex N=2 string
\cite{GSW}. The general solution $ x_0(t,\sigma)
= x_L(\t) + x_R(\bar\t) $ is a sum of the analytic and
anti-analytic modes $x_L(\t), \ x_R(\bar\t),$ which are not necessarily
complex conjugate. For each real point $x^\m ,$ the
parameters $\t=\t_L$ and $\bar \t=\t_R $ should be determined by a complex
retarded-time construction, a complex analog of the real one.
Complex source of the real KN solution
corresponds to two complex conjugate \emph{straight}
world-lines,(\ref{CWL}) and (\ref{CWR}). The
 complex light cone (\ref{KL}) is split into the Left and Right
complex null planes. In accord to the retarded-(advanced)-time equations  $\t^\mp = t \mp \tilde r ,$  intersections of the Left null planes with Left CWL, together with the conjugate Right structure,  determine four retarded-advanced complex time parameters
\cite{BurKerr,BurNst} \bea \t_L^\mp &=&
t \mp (r_L + ia\cos\theta_L) \label{Lretadv} \\
\t_R^\mp &=& t \mp (r_R + ia\cos\theta_R) \label{Rretadv}. \eea
\begin{figure}[ht]
\centerline{\epsfig{figure=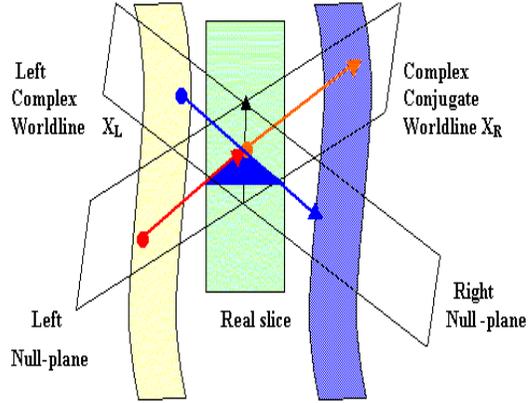,height=6cm,width=8.5cm}}
\caption{The complex conjugate Left and Right null planes generate four roots for
the Left and Right retarded and advanced times.}
\end{figure}
 The real slice of Kerr geometry is formed by the complex conjugate roots
$\t_L = t_L +i\sigma_L$ and $\t_R = t_R +i\sigma_R ,$ which sets the constraint
$\sigma_L = - \sigma_R =\sigma=a\cos \theta $ resulting in dependence of the
angular directions  of the null rays $\theta$ on the world-sheet parameter
$\sigma.$ Since $|\cos \theta| \le 1 ,$ we obtain $\sigma \in [-a, a] ,$ and consequently,
\emph{the complex string should have the endpoints $\sigma = \pm a ,$} and
is to be open.  The complex endpoints are mapped to  the  `north' and `south' lines of the  Kerr congruence, corresponding to  $\theta =0,\pi ,$ as it is shown in Figure 3.

\paragraph{World sheet orientifold.} The complex open string boundary
conditions require the \emph{orientifold}
structure \cite{BurCStr,GSW,BBS} which
 turns the open string in a closed but
folded one. The world-sheet parity transformation $ \sigma
\to - \sigma $ reverses orientation of the world sheet, and covers
it second time in mirror direction. Simultaneously, the Left and
Right modes are to be exchanged. Two oriented copies of the interval $\Sigma = [-a, a] ,$
$\Sigma^+ = [-a, a],$ and $ \Sigma^- = [-a, a]$ are joined,
forming a
 circle $ S^1 = \Sigma ^+\bigcup \Sigma ^-
,$ parametrized by $\theta ,$ and the map $\theta \to \sigma=a\cos
\theta $ covers the world-sheet twice.
 The string modes $x_L(\t), \quad x_R(\bar\t),$ are extended on
the second half-cycle by the well known extrapolation,
\cite{GSW,BBS} \be x_L(\t^+) = x_R(\t^-); \quad x_R(\t^+) =
x_L(\t^-) , \label{orbi}\ee which forms the folded string, in which
the retarded and advanced modes are exchanged every half-cycle.
Revers of $\sigma$ acts on the complex radial
coordinate $\tilde r = r+ia \cos \theta$ as anti-analytic involution
\be C: r +ia \cos \theta  \to r -ia \cos \theta .
\label{Ctr} \ee Another anti-analytic involution is reflection of the
real radial coordinate, $R: r \to - r .$ Analyticity of the
world sheet is restored by use of double anti-analytic involution ${\cal T}=CR .$
Applying ${\cal T}\tilde r = -\tilde r $ to the Left CWL, we obtain
\be  {\cal T}: \ \t_L^\pm \to \t_L^\mp \label{cT} ,\ee which sets orientifold parity
between retarded and advanced times of the same Left CWL allowing to leave aside the
anti-analytical Right structure.
\begin{figure}[ht]
\centerline{\epsfig{figure=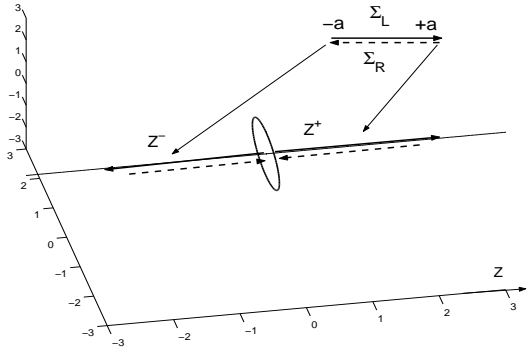,height=5cm,width=7cm}}
\caption{Ends of the open complex string, associated with quantum
numbers of quark-antiquark pair, are mapped onto the real
half-infinite $z^+, z^-$ axial strings. Dotted lines indicate
orientifold projection.}
\end{figure}
\section{Embedding of the complex string in the real Kerr-Schild geometry}
 The complex world-sheet parameters $\t^\pm =t \pm r + ia \cos \theta$ are mapped in
 the real KS geometry as coordinates $\rho^\pm = t \pm r$ and  $\theta ,$ and therefore,
 the Kerr coordinate transformations (\ref{oblate}) may be considered as \emph{embedding}
 of the world-sheet in the bulk of the \emph{real} KS geometry. Simultaneously, it is the
 embedding of the twistorial coordinates of the adjoined Left null planes. The first twistor coordinate $Y=\psi_L^1/\psi_L^0$ determines directions of the null rays $Y=e^{i\phi} \tan \frac \theta 2 $ in accord to (\ref{e3}).  The coordinate $\phi $ adds extra dimension to the world-sheet, extending it to a membrane formed by merging of the real closed Kerr string with the complex open string, similar to the model of string/M-theory unification \cite{BBS,BurAlter}.

 The given by Kerr theorem two solutions $Y^\pm(x)$ have different embeddings in the real KS geometry. For simplicity, we consider the Kerr
source at rest in the standard orientation of angular momentum along the
z-axis. Assuming that the complex source is positioned at the point $\vec
x_L=(0,0,-ia),$ and setting $x_0^\m (0)=0$ and $u^\m=(1,0,0,0)$ in (\ref{CWL}),
  we rewrite the Left CWL and its time-derivatives in the null coordinates
  (\ref{nullcoord}) as follows,
\be x_L (\t_L)=\{u_L,v_L,\z _L,\Z _L\}=2^{-\frac 1 2}\{(-ia +\t _L),(-ia
- \t _L), \ 0,\ 0 \}\label{XLna}, \ee
\be \dot x_L=\{\dot u_L,\dot v_L, \dot \z _L, \dot \Z _L\} =
2^{-\frac 1 2}\{1, -1, \ 0, \ 0 \} \label{dotnul} .\ee
Coefficients $A, \ B, \ C $ were obtained for this case in \cite{BurKerr,BurNst},
and take the form
 \bea
 A_L &=& (\Z  - \Z_L(\t)) \`v_L(\t) - (v-v_L(\t)) \`\Z_L(\t) ;\nonumber\\
 B_L &=& (u-u_L(\t)) \`v_L(\t) + (\z - \z_L(\t) )\`\Z_L(\t)
  - (\Z - \Z_L(\t)) \`\z_L(\t) - (v - v_L(\t)) \`u_L(\t) ;\nonumber\\
C_L &=& (\z - \z_L(\t) ) \`u_L(\t) - (u -u_L(\t)) \`\z_L(\t), \label{ABC} \eea
which yields for the Left CWL the expressions
\be A_L= - \frac 12 (x-i y), \quad B_L = -z - ia , \quad C_L= \frac
12 (x+iy) \label{ABCL} ,\ee
and using generating function $F_L(T^A) =A_L Y^2 + B_L Y +C_L $ we obtain
\be Y_L^\pm =\frac { - z -ia \pm \tilde r_L}{x-i y} \label{YpmL} ,
\ee
where
\be \tilde r_L = (B_L^2 -4 A_L C_L)^{\frac 12} = [x^2 +y^2 + (z+i
a)^2]^{\frac 12} \label{tra1}\ee  corresponds to the complex radial distance
of the Appel source. We will further omit the
 suffix $L ,$ if we work only with the Left CWL.
Representing the complex radial distance in the form $\tilde r=r
+i\eta $ and squaring (\ref{tra1}), we obtain two real
equations. The first of them, $r\eta = az ,$ may be treated as an analog of the
relation $ z=r\cos\theta ,$ and, setting $\eta=a\cos\theta ,$ we obtain
\be \tilde r = r+ia \cos\theta, \quad z=r \cos\theta . \label{trz}
\ee
The second equation, following from (\ref{tra1}), may be combined
with (\ref{trz}) and yields
$ x^2+y^2 = (r^2 + a^2)\sin^2\theta ,$ which may be split into
four different linear relations
$ x+iy = (r \pm ia)e^{\pm i\phi}\sin\theta ,$ characterized by
different combinations of the signs $\pm i a$ and $\pm i\phi .$
We obtain four sheets corresponding to four different embeddings of the
world-sheet of the complex string.

\paragraph{Orientifolding the bulk of the Kerr space.} Embedding of the world-sheet fixes also the map of the Left null planes into the real rays of the Kerr congruence
in accord to (\ref{e3}).
Considering the  Kerr coordinate relations (\ref{oblate}) as \emph{analytic} one,
we can represent the given by (\ref{YpmL}) solution $Y^+ $   in  the form
\be Y^+ =\frac {e^{i\phi} (1 -  \cos \theta)}{\sin\theta} =
e^{i\phi}\tan\frac \theta 2  ,\ee which is just the original form of $Y$
given in \cite{DKS}.
In the same time, for $Y^-$ we obtain a bizarre expression
$ Y^- =\frac { - (r +ia)(1+ \cos \theta)}{x-i y} ,$ which shows that `analytic' embedding (\ref{oblate}) is inconsistent with $Y^- ,$ and consistency is related with  `anti-analytic' transformation
\be x+iy = (r-ia)e^{i\phi}\sin\theta, \ z= r\cos \theta ,\label{xyzma}\ee which results in
the relation
\be Y^- =\frac { - e^{i\phi}(1+ \cos \theta)}{\sin\theta}
=-e^{i\phi} \cot \frac \theta 2 = - 1 /\Y^+ \label{YmL} . \ee
We reveal that the solutions  $Y^+$ and $Y^-$ are related by antipodal involution
\be I^\star : \ Y^+ \to Y^-= - 1 /\Y^+ ,
\label{IYApod} \ee
which is correlated with the crossing-symmetry of the Left and Right  world-sheet structures
(\ref{orbi}), as it is shown in Fig.2.
Antipodal involution
\be I^\ast: \ \theta \longrightarrow \tilde {\theta} = \pi
-\theta, \quad \phi \longrightarrow\tilde\phi =\phi +\pi
\label{apodphth}\ee
was studied for the case of non-rotating black
holes in \cite{ChaGib}, where authors considered formation of the non-orientable or single-exterior BH  solutions.

\paragraph{Calabi-Yau twofold from the Kerr theorem.}
As follows from (\ref{YpmL}), coordinates  $x^\m_L(\t^-)$ and 4-velocity $u^\m(\t^-)$ of the
Left CWL determined by the Left complex retarded time $\t^-_L$ lead to two solutions
$Y^\pm (x^\m, \t^-_L ).$
One sees from (\ref{Ypm}) that $Y^+$ and $Y^-$ differ only in the sign of
$\tilde r .$  In accord to (\ref{cT}), this difference corresponds to change the Left
retarded time $\t^-_L$ to the Left advanced time  $\t^+_L ,$
and therefore, the second solution should be identified with the
Left advanced one $Y^- (x^\m, \t^+_L ).$
The both Left solutions are analytic, since the transfer to $Y^- (x^\m, \t^+_L )$ may be represented
as result of two anti-analytic transformations: complex conjugation $C$ and antipodal map
$ I^\star ,$ like the two-step transformation of the world-sheet (\ref{cT}).
On the real slice, the Left and Right retarded solutions, $ Y^+ (x^\m , \t^-_L)$ and
$\bar Y^+ (x^\m , \t^-_R) ,$ are complex conjugate, as well as the corresponding retarded
times $\t^-_R$ and $\t^-_L .$
Antipodal map (\ref{IYApod}) connects the Left advanced solution $Y^- (x^\m , \t^+_L)$
with the Right retarded one $\bar Y^+ (x^\m, \t^-_R) ,$ forming a crossing-symmetry, as shown
in Fig.2. On the real slice the Left retarded and advanced solutions are rigidly
correlated. However, this correlation is destroyed when $x^\m$ escapes the real slice.
Complex analyticity of the Kerr theorem is used by integration of the field equation \cite{DKS}, which  assumes independence of the Left generating functions $F_{ret} \equiv F \{T^A, \t^+ \} $ and $F_{adv} \equiv F \{T^A, \t^- \} $  and the corresponding solutions
$Y_{ret}^-(x^\m) \equiv Y^-(x^\m, \t^-_L)$ and $Y_{adv}^+ \equiv Y^+(x^\m , \t^+_L) .$

Each of the functions $F_{ret}\{T^A  \} $ and $F_{adv}\{T^A, \} $ creates two sheets of
the Kerr-Schild geometry. The space-time, determined by two independent generating function
turns out to be four-sheeted and should be determined by multi-particle version of the Kerr
theorem considered in \cite{Multiks,MultPart}.
In particular, the analytic bipartite generating function $F_2(T^A)$ should has the form
$ F_2(T^A) = F_{ret} (T^A ) F_{adv} (T^A ) .$ As a consequence of (\ref{FYYpm}),  \be F_2(T^A)=
A_{ret}A_{adv}(Y-Y_{ret}^+(x^\m))(Y-Y_{ret}^-(x^\m))(Y-Y_{adv}^+(x^\m))
(Y-Y_{adv}^-(x^\m))  \label{F4Y} ,\ee and the equation (\ref{F0KerrTeor})
acquires the fourth degree in $Y $ forming  \emph{a quartic in $CP^3 $}, which is
a canonic form of the Calabi-Yau two-fold, \cite{BBS,GSW}.
\begin{figure}[h]
\begin{minipage}{14pc}
\includegraphics[width=14pc]{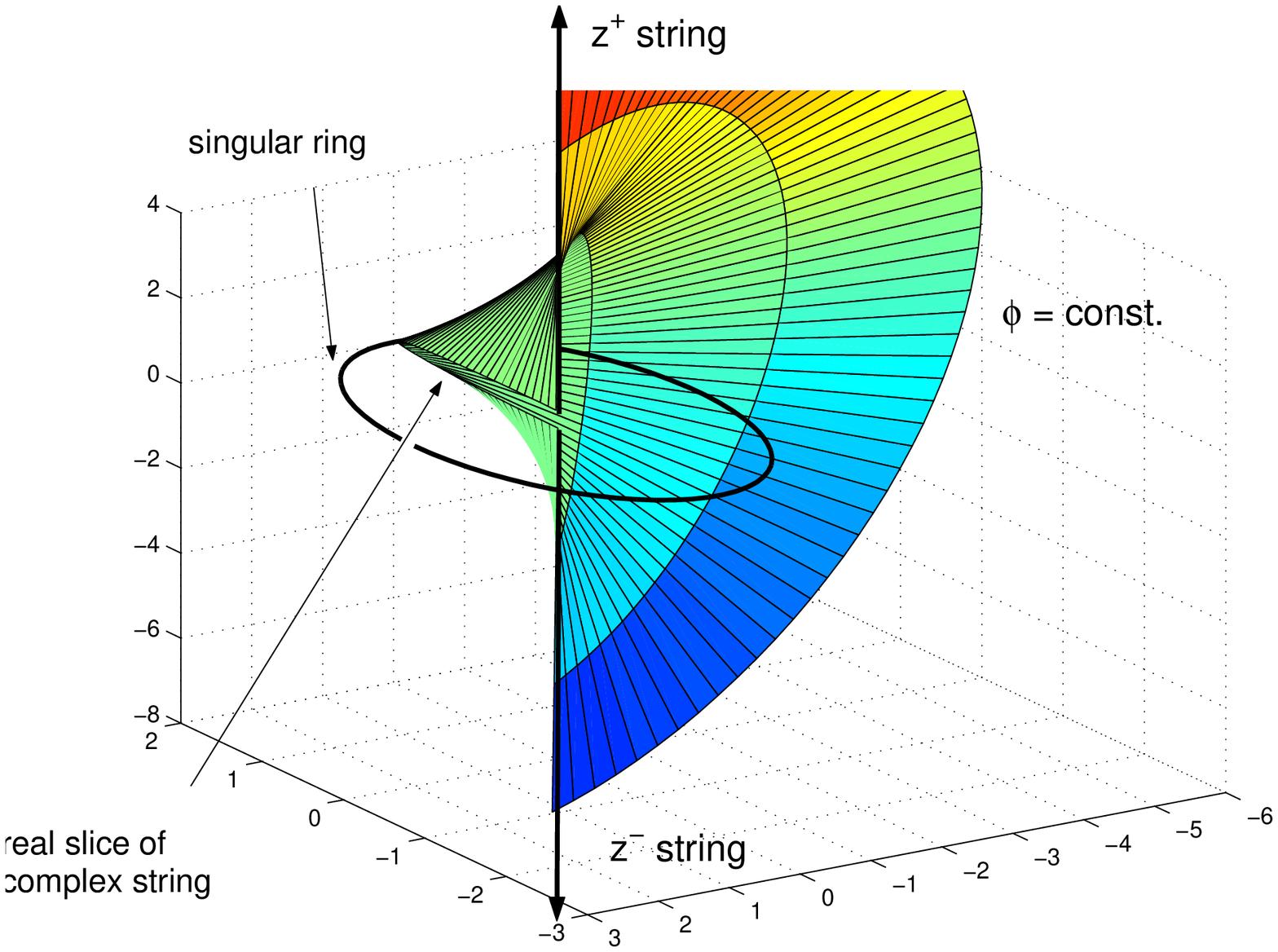}
\caption{\label{label} One sheet of the K3 for $r>0$ and $\phi=const.$ Kerr congruence
is tangent to singular ring at $\theta=\pi /2 .$}
\end{minipage}\hspace{2pc}%
\begin{minipage}{12pc}
\includegraphics[width=12pc]{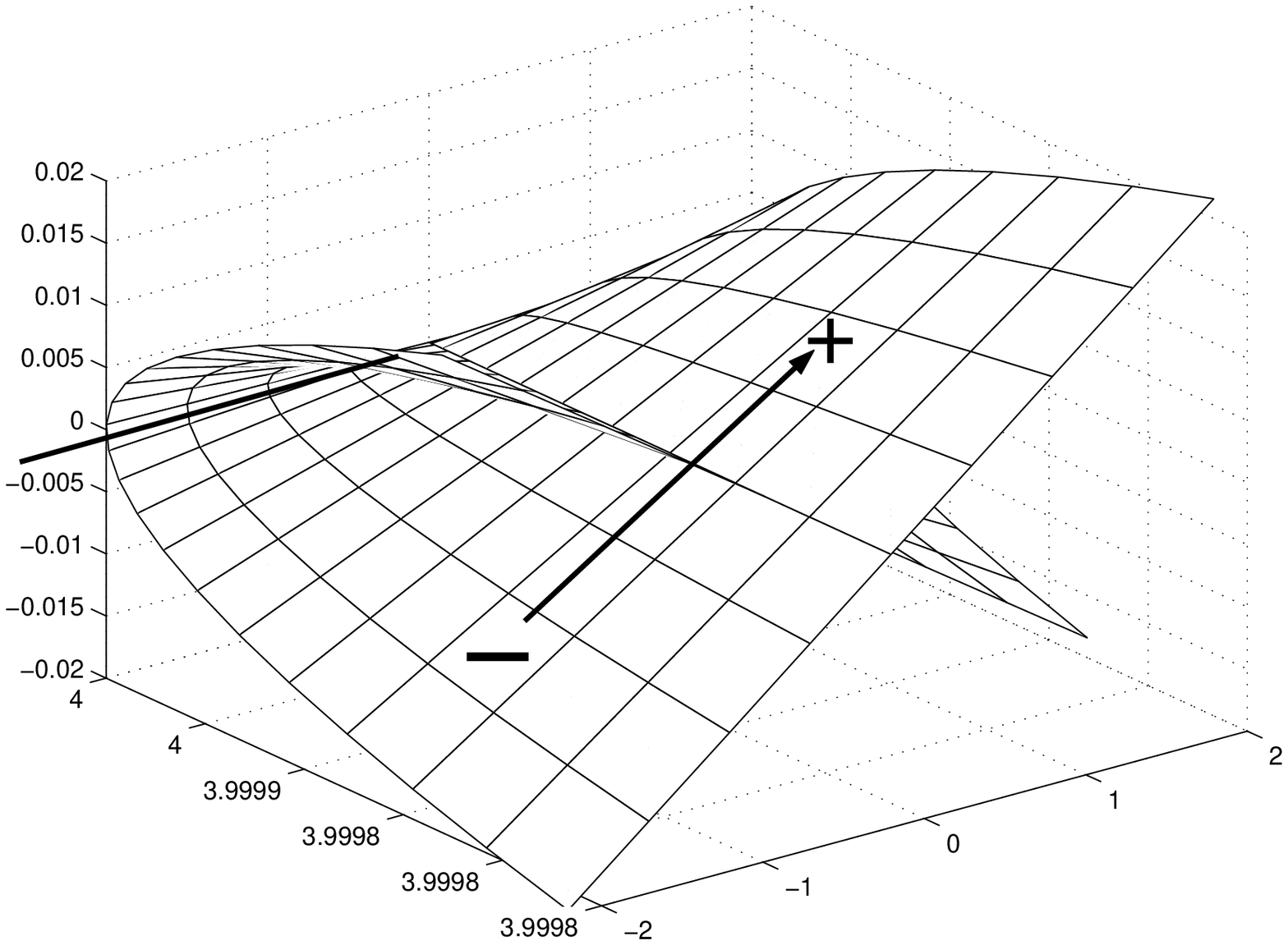}
\caption{Section of the K3  near the Kerr ring corresponding to $\phi=const.$.
Two sheets ($r>0$ and $r<0$) form a M\"obius strip covering the space-time twice.}
\end{minipage}
\end{figure}
For each point $x^\m \in M^4 ,$ solutions  $Y(x^\m)$ may be transformed to twistor
coordinates $T^A $  by means of (\ref{TA}) and
(\ref{nullcoord}).

The four solutions $Y(x^\m)$ correspond to different sheets of the
K3 surface and perform different embeddings of the Kerr congruence in the real Kerr
geometry, in accord to (\ref{e3}).
Antipodal map acts on the real null directions $e^{3\m}$
as inversion of its spacelike part, and therefore, the congruences generated by  functions
 $F_{ret}$ and $F_{adv}$ are not to be equivalent.
 In particular, the tangent to Kerr congruence normalized null vector
\be k^\m (x) = e^{3\m}(x)/|e^{30}|,
\label{kmu}\ee is transformed as follows
 \be I^\star  : \  k^\m_{out}=\{ k^0, \vec k \} \to k^\m_{in} =\{ k^0, - \vec k \} .
 \label{Kk}\ee

A slice of the K3 surface corresponding to $\phi=const.$ and $r>0 $ is presented
in Fig.4. One see that this sheet is a ruled surface generated by twistor null lines parametrized
by affine parameter $r\in [0, \infty] .$ By variation of $\theta \in [0,\phi] ,$ the lines
rotate on angle $\theta .$ The lines of the equatorial plane, $\theta=\pi/2 ,$ are tangent to
the Kerr ring.  Analytic extension of this surface $\phi=const.$ to region of negative
$r \in [0, -\infty] ,$ as shown in Fig.1 and  Fig.5, gives the second sheet of the K3
which corresponds to `negative' sheet of the Kerr geometry.
The surface form  a M\"obius strip, since the boundary lines at $\theta=0$ and $\theta=\pi $
have opposite orientation.
  Antipodal map $I^\star$ creates two extra sheets of the K3 which are advanced folds of the
  Kerr congruence. The global antipodal involution (\ref{apodphth}) transforms the
  surface $\phi=const.$ into the surface $\phi+\pi=const.$ with simultaneous involution
  of the null generators. Although the surfaces $\phi=const.$ and $\phi+\pi=const.$ are smoothly
  matched, the North and South boundary lines of these surfaces have opposite orientation.
So far as the orientation is changed by the transfer from $r>0$ to $r <0 ,$ one sees that
analytic gluing of the four sheets of the K3 are to be reached under a cross-symmetric matching
of the boundaries of the retarded and advanced folds of the Kerr congruence.
The four folds of the K3 surface form orientifold parity of the Kerr geometry, under which
the principal null directions $e^3$ are doubled by the joined in-going and out-going null rays.

\noindent\emph{String excitations.}
The Killing direction $K^\m $ may be expressed  in  Cartesian coordinates via coordinates of the CWL
 \be K^\m  = \d_t x^\m_L(\t),\label{Kmu}\ee
and the coefficients $A,B,C$ (\ref{ABCL}) turn out to be functions of the
coordinates and 4-velocity  of the CWL $u^\m .$
The string excitations are related with radiation and a recoil leading to non-stationarity of the solution.
The exact non-stationary KS solutions are unknown, however the recoilless  solutions were obtained in \cite{BurA,BurPreQ}.
 A recoil assumes a deviation of the 4-velocity   $\delta\`x^\m_L(\t)|_{\t^-}$ for the moments $t > 0 ,$ which
 creates a difference between parameters $A,B,C,$ determined for the retarded and advanced times $\t^- $ and $\t^+ ,$
resulting in two independent generating functions $F_{ret}(T^A)$ and $F_{adv}(T^A).$ This duplications of the Kerr congruence generates the four sheets of the K3 surface. Independence of the retarded and advanced solutions creates
breakdown of the orientifold parity of the Kerr congruence. The North and South boundaries of the Kerr sheets are decoupled,  but the smooth pairwise matching of the four sheets is retained, resulting in analyticity of the K3 surface.

\section{Outlook.}

It has been discussed in I that the Kerr-Schild geometry displays
striking parallelism with basic structures of the superstring theory.
However, along with wonderful parallelism, the KN geometry displays also
several peculiarities.

The principal of them is that the Kerr real and complex Kerr strings, and the related
stringy structures such as orientifold and formation of the M2-brane, are
living in the four-dimensional space-time. There appears a
wonderful possibility that compactification of higher dimensions
may be replaced by complexification.

Second peculiarity is related with  characteristic parameter of the Kerr
strings $a=\hbar/m ,$ which corresponds to Compton scale, and this new
parameter  makes this version of the superstring theory to be closer to
particle physics.

Finally, we notice that among the consistent critical strings in dimensions
d=26 and d=10 , there is also the
complex N=2 string \cite{GSW,Gibb}, which has the real critical dimension four and
could be used as a basis of some four-dimensional string theory, \cite{DAddaLiz}.
The principal obstacle for its application emerges from its signature, which may only
be (2,2) or (4,0), which conflicted with its embedding in the real minkowskian
space-time. Up to our knowledge, this trouble was not resolved so far,
and the initially enormous interest to this string seems to be dampened.
We note that  N=2 string has organic embedding in
 the complexified 4d Kerr geometry.
 Hermitian action for the Kerr's complex  world line (CWL),
 $ S=- \frac 1{2\pi} \int d \t d \bar \t g_\mn{\d_\t x^\m \d_{\bar\t} \bar x^\n} ,$
 corresponds to bosonic part of the N=2 string action \cite{OogVaf,GSW,Gibb}.
 Generalization $\t$ to complex super-time $\cal T$ turns CWL into super-world-line
$ X^\m (\cal T),$
 \cite{BurCStr,BurSPStr}, and leads to action of the N=2 superstring
 \cite{GSW}.
 The extra world-sheet spinor field $\psi(\t)$ corresponds
 to the Left null planes, generators of the K3 and Kerr congruence.
Detailed treatment of this structure goes
 beyond the frame of this paper and will be given elsewhere.

\section*{Acknowledgments}
This work is supported under the RFBR grant 13-01-00602. Author thanks Theo M. Nieuwenhuizen
for permanent interest to this work and useful conversations, and also thanks
Dirk Bouwmeester for invitation to Leiden University, where these ideas were discussed
in details with him and members of his group: Jan W. Dalhuisen,
V.A.L. Thompson and J.M.S. Swearngin. Author is also thankful to Laur Jh\"arv for useful
discussion at the Tallinn conference "3Quatum" and given reference to the paper \cite{LaJoh}.

\end{document}